\documentclass[11pt]{article}
\usepackage{amsmath,amssymb}
\newcommand{\beq}{\begin{equation}}
\newcommand{\eeq}{\end{equation}}
\newcommand{\ba}{\begin{array}}
\newcommand{\ea}{\end{array}}
\newcommand{\bea}{\begin{eqnarray}}
\newcommand{\eea}{\end{eqnarray}}
\newcommand{\bean}{\begin{eqnarray*}}
\newcommand{\eean}{\end{eqnarray*}}

\newtheorem{theorem}{Theorem}[section]
\newtheorem{prop}[theorem]{Proposition}
\newtheorem{lem}[theorem]{Lemma}
\newtheorem{defi}[theorem]{Definition}
\newtheorem{remark}[theorem]{Remark}

\newtheorem{proof}{Proof.}
\newenvironment{pf}{\begin{proof} \rm}{\hfill$\square$ \end{proof}}
\newcommand{\CH}{{\cal H}}
\newcommand{\CP}{{\cal P}}
\newcommand{\CD}{{\cal D}}

\newcommand{\CW}{{\cal W}}
\newcommand{\CZ}{{\cal Z}}
\newcommand{\CS}{{\cal S}}

\newcommand{\CN}{{\cal N}}
\newcommand{\CV}{{\cal V}}
\newcommand{\CM}{{\cal M}}
\newcommand{\CQ}{{\cal Q}}

\makeatletter
\@addtoreset{equation}{section}

\makeatother

\newcommand{\NN}{{\mathbb N}}

\def\al{\alpha}

\newcommand{\cmp}[3]{Comm. Math. Phys. {\bf #1} (#2), #3}

\newcommand{\faa}[3]{Funct. Anal. Appl. {\bf #1} (#2), #3}

\newcommand{\lmp}[3]{Lett. Math. Phys. {\bf #1} (#2), #3}
\newcommand{\jgp}[3]{J. Geom. Phys. {\bf #1} (#2), #3}

\newcommand{\rmp}[3]{Rev. Math. Phys. {\bf #1} (#2), #3}

\newcommand{\jmp}[3]{Jour. Math. Phys. {\bf #1} (#2), #3}
\newcommand{\rref}[1]{(\ref{#1})} 

\def\dsl{\displaystyle}

\newcommand{\del}{{\partial}}

\def\dpt#1#2{\frac{\partial #1}{\partial t_{#2}}}

\def\H#1{H^{(#1)}}

\def\Mfive{\CM_{3}^{(5)}}

\def\alg{{\mathfrak g}}

\newcommand{\fraksl}{{\mathfrak s}{\mathfrak l}}
\newcommand{\CHp}{{\CH_+}}

\def\var{manifold}

\def\bih{bihamiltonian}
\def\varb{\bih\ \var}

\def\varb{\bih\ \var}

\def\V#1{{\CV^{(#1)}}}
\begin{document}
\begin{titlepage}
\begin{flushright}
Ref. SISSA 123/98/FM
\end{flushright}
\vspace{1truecm}
\begin{center}
{\Huge
Reduction of bihamiltonian systems  and separation  of variables:
an example from the Boussinesq hierarchy}
\end{center}
\vspace{0.5truecm}
\begin{center}
{\Large
G. Falqui${}^1$,
F. Magri${}^2$, and
G. Tondo${}^3$}\\
\vspace{.5truecm}
${}^1$ SISSA, Via Beirut 2/4, I-34014 Trieste, Italy\\
E--mail: falqui@sissa.it\\
${}^3$ Dipartimento di Matematica, Universit\`a di Milano\\
Via C. Saldini 50, I-20133 Milano, Italy\\
E--mail: magri@vmimat.mat.unimi.it\\
${}^3$ Dipartimento di Scienze Matematiche,
Universit\`a di Trieste\\
Piazzale Europa 1, I-34127 Trieste, Italy\\
E--mail: tondo@univ.trieste.it
\end{center}
\vspace{1.truecm}
\abstract{
We discuss the
Boussinesq system with $t_5$ stationary,
within a general framework for
the analysis of stationary flows of $n$--Gel'fand
Dickey hierarchies.
We show how a careful use of its bihamiltonian
structure can be used to provide a set of separation 
coordinates for the
corresponding Hamilton--Jacobi equations.
}
\\
\vspace{1truecm}
\begin{center}
Work supported by 
the G.N.F.M. of the Italian C.N.R.
\end{center}
\end{titlepage}
\setcounter{footnote}{0}

\section{Introduction}\label{sec1}
It is well known that the Boussinesq equation 
$ u_{tt}=\frac{1}{3}(-u_{xxxx}+4 u_x^2+4 u u_{xx})$
represented as a first order system of equations
in the variables $(u,v)$
\begin{equation}\label{eqbo2}
\begin{split}
u_{{t_{{2}}}}&=-u_{{xx}}+2\,v_{{x}}\\
v_{{t_{{2}}}}&=v_{{xx}}-2/3\,u_{{xxx}}+2/3\,uu_{{x}}
\end{split}
\end{equation}
is a member of an infinite
hierarchy of evolution equations. These equations are defined as follows.
In the space $\CM_3$ of the third--order differential operators
$L=\partial_x^3 +u(x)\partial_x +v(x)$, one considers the third 
root $L^{\frac{1}{3}}$ of $L$, i.e. a pseudodifferential
operator of the form
\begin{equation}
L^{\frac{1}{3}}=\del+\sum_{i=1}^\infty q_i(u,v)\del^{-i}
\end{equation}
satisfying $(L^{\frac{1}{3}})^3=L$.
By means of the powers of such a third root, one defines an infinite family
of commutative flows on $\CM_3$ as
\begin{equation}\label{Bouflow}
\dpt{}{p} L = [L, (L^{\frac{p}{3}})_+]
\end{equation}
where $(\cdot)_+$ is the projection on the purely differential part of a
pseudodifferential operator.
{\em Stationary flows} for such a system correspond to that
subspace of operators $L$ satisfying the (non--linear) relation
$[L, L^{\frac{p}{3}}]=0$ for some $p$.\\
In this paper we will discuss the
case $p=5$.
The corresponding equation of the hierarchy reads:
\begin{equation}
\label{eqbo5}
\begin{split}
u_{{t_{{5}}}}&
=-1/9\,u_{xxxxx}+5/9\,u_{{x}}u_{{xx}}+5/9\,uu_{{xxx}}+5/3\,u_{{xx}}v+\\&
5/3\,u_{{x}}v_{{x}}-5/9\,{u}^{2}u_{{x}}-10/3\,vv_{{x}} \\
v_{{t_{{5}}}}&=-1/9\,v_{xxxxx}+{10}/{9}
\,u_{{xxx}}v+20/9\,u_{{xx}}v_{{x}}+5/3\,u_{{x}}v_{{xx}}+\\
& 5/9\,uv_{{xxx}}+-5/3\,{v_{{x}}}^{2}-
5/3\,vv_{{xx}}-10/9\,uvu_{{x}}-5/9\,{u}^{2}v_{{x}},
\end{split}
\end{equation}
and one can see that the space $\Mfive$ of the zeroes of
$\dsl\dpt{}{5}$ can parametrized by the space of the Cauchy data
of two fifth order ODE in two variables.
On such a ten--dimensional manifold we
consider the (restriction of the)
first {\em four} non--trivial flows of the Boussinesq hierarchy
\begin{equation}\label{4flows}
\dpt{}{p} L = [L, (L^{\frac{p}{3}})_+],\quad p=1,2,4,7.
\end{equation}
Our aims are to study the bihamiltonian structure
of these equations,
and to show how a careful use  of such structures
explicitly provides  coordinates in which the corresponding
Hamilton--Jacobi equations are separable.\par
In the first part we show that on
$\Mfive$ one can construct a linear pencil of Poisson brackets
$\{F,G\}_\lambda =\{F,G\}_1 -\lambda \{F,G\}_0$ with the following
properties:
\begin{enumerate}
\item
they have two Casimir functions $ H(\lambda )= H_1\, \lambda+ H_2$ and
$K(\lambda) =H_3\, \lambda^3 +H_4\, \lambda^2 + H_5\,\lambda +H_6$;
the functions $H_1$ and $H_3$ are Casimir functions of $\{F,G\}_0$,
 the functions $H_2$ and $H_6$ are Casimir functions of
$\{F,G\}_1$;
\item
the functions $H_2$,$H_4$,$H_5$ and $H_6$ are the Hamiltonian functions,
with respect to the  Poisson bracket
$\{F,G\}_0$, of the four vector fields under scrutiny, while
the functions $H_1$, $H_3$, $H_4$, $H_5$ are their Hamiltonian functions 
w.r.t  the  Poisson bracket $\{F,G\}_1$;
\item
all these functions are in involution  w.r.t.
the whole Poisson pencil.
\end{enumerate}
Then we prove that the  bihamiltonian structure of the
vector fields~\rref{4flows} can be used to
integrate the equations of motion according to the following two--step 
scheme.
\par
At first one considers a level surface $S_0$ of the functions $H_1$ and
$H_3$. It is an eight--dimensional symplectic leaf of the Poisson bracket
$\{F,G\}_0$. The given vector fields are tangent to $S_0$, so they can
be restricted to this manifold. Then one remarks that both Poisson
brackets $\{F,G\}_0$ and $\{F,G\}_1$ reduce to $S_0$ by a
Marsden--Ratiu reduction procedure. So one can conclude that  $S_0$ inherits
a Poisson--Nijenhuis structure, i.e., $S_0$ is a manifold endowed with a
symplectic form $\omega$ (induced by $\{F,G\}_0$) and with a
compatible Nijenhuis tensor $N$ (induced by $\{F,G\}_1$).\par
The second step is to construct a set of separation
variables defined on $S_0$.  To this end, we shall use  a kind of  coordinates
introduced in~\cite{MaMar} under the name of {\em
Darboux--Nijenhuis} coordinates. 
They were exploited in~\cite{MT}, as separation variables for the
H\'enon--Heiles type systems obtained from the stationary  
flows of the KdV hierarchy.  In this
paper, we explicitly compute such coordinates for the stationary Boussinesq
flows, and finally we show  that they are separation  variables
for the Hamilton--Jacobi equations associated with the Hamiltonian functions
$(H_2, H_4, H_5, H_6)$.
\par
In our opinion these flows provide a good example of the
tight connection between separable coordinates and the bihamiltonian
structure of soliton equations. We have chosen to study in detail
this problem, in order to show that these methods can be effectively
used to treat systems that are more involved than the  
KdV cases (see, e.g.,~\cite{Alber,AFW,BoNo,DKN}),
which fall into the class
of \varb\ of {\em maximal rank} discussed in~\cite{GZ93}.
Accordingly, we will illustrate and display the appropriate
computations, rather than proving general propositions.
However, we would like to point out that these kind of results
are not specific of the Boussinesq equations. Indeed, as it will
be clear from our arguments, they have analogues holding,
{\em mutatis mutandis},
for a wide class of stationary flows of $n$--Gel'fand--Dickey hierarchies.
Such a formalization will be the subject of a further publication.

\section{The Central System and its reduction}\label{sec2}
To obtain Lax pairs and Hamiltonian structures for the
$t_5$ stationary Boussinesq system (hereinafter the {\em Bsq$_5$} system),
we recall the set up for the KP theory
discussed in~\cite{cfmp5,fmp1}.

We consider the space $\CM$ of sequences of Laurent series
$\{\H{k}\}_{k=1\ldots\infty}$ (the {\em currents} of the theory)
having the form
\begin{equation}
\H{0}=1,\qquad
H^{(k)}=z^k+\sum_{l\ge 1}H^k_l z^{-l},
\end{equation}
where  $H^k_l$ are scalars.
On $\CM$ we define a family of vector fields
as follows. We associate with a point $\{\H{k}\}_{k=0,\dots,\infty}$ in $\CM$
the linear span $\CHp$ of  the currents
$\H{k}$:
\begin{equation}
\CHp=\langle 1, \H{1},\H{2}\ldots \rangle,
\end{equation}
which is a subspace in the vector space $\CH$ of 
Laurent series in the formal variable $z$.
Then we consider the following invariance relations:
\begin{equation}
\left(\dpt{}{j}+\H{j}\right)\CHp\subset\CHp,
\end{equation}
as the {\em defining equation} for the $j^{th}$ vector field of the family.\\
Explicitly we have:
\begin{equation}
\label{CS}
\frac{\partial H^{(k)}}{\partial t_j}=H^{(j+k)}-H^{(j)}H^{(k)}+\sum_{l=
1}^kH^j_lH^{(k-l)}+
\sum_{l=1}^jH^k_lH^{(j-l)}.
\end{equation}
which will be called in the sequel
the {\em Central System}.
It has the following
properties:
\begin{itemize}
\item\label{commu} commutativity: $\dsl{[\dpt{}{j},\dpt{}{k}]=0}$;
\item\label{exa} ``exactness'': $\dsl{
\dpt{}{k}\H{j}=\dpt{}{j}\H{k}}.$
\end{itemize}

The connection of the Central System
with the usual formulation of the KP theory (see, e.g.~\cite{SS, DikBook} and
the references quoted therein) as a system of Lax
evolution equations for the Sato pseudodifferential operator
$\CQ=\del+\sum_{i=1}^\infty q_i \del^{-i}$, as well as the proof of the two
properties stated above, can be found in~\cite{cfmp5}.
Here we simply recall that  reductions of KP to the
$n$ Gel'fand--Dickey case are obtained
via the constraint
\begin{equation}\label{hnzn}
\H{n}\equiv z^n
\end{equation}
a requirement
equivalent to the usual constraint $\left(\CQ^n\right)_-=0$.
In particular
the Boussinesq hierarchy
corresponds to the case $n=3$.

It should be noticed that on the subset
$\CS_n$ of $\CM$ formed by those points satisfying~\rref{hnzn}
we can read the $n^{\mbox{th}}$ equation of CS as the constraint equation
\begin{equation}\label{constr}
\H{j+n}=z^n\H{j}-\sum_{l=1}^nH^j_l\H{n-l},
\end{equation}
which allows to recursively define the Laurent coefficients of $\H{k}, k>n$,
in terms of the coefficients of the first $n$ currents.
More formally:
\begin{prop}\label{invsub}
The submanifold $\CS_n$ is the subset of $\CHp$
given by the equation
\begin{equation}
z^n(\CHp)\subset \CHp,\label{zinv}
\end{equation}
i.e., the set of the points where the operator
of multiplication by $\lambda=z^n$ leaves the space
$\CHp$ invariant.
Hence, $\CS_n$ is generated by the first $n$ currents
\begin{equation}
\H{0}\equiv 1,\H{1},\ldots,\H{n-1}
\end{equation}
over the space of polynomials in $\lambda$ with coefficients
in the set
\begin{equation}
\{ H^1_{l_1},H^2_{l_2},\ldots, H^{n-1}_{l_{n-1}}\}, l_i\in \NN.
\end{equation}
The reduced equation of motion, CS${}_n$ are simply the {\em restriction} of
CS to $\CS_n$.
\end{prop}
In the case $n=3$, thanks to Proposition~\ref{invsub}
the degrees of freedom are thus collected into the two series
\begin{equation}
\H{1}\equiv h = z+\sum_{i=1}^\infty h_i z^{-i}\ \mbox{ and }
\H{2}\equiv k = z^2+\sum_{i=1}^\infty k_i z^{-i}.
\end{equation}
At this level, we have still a dynamical system with an infinite number of
degrees of freedom.
To obtain finite--dimensional dynamical systems representing the restriction
of the flows to the stationary manifolds, one can remark that
the set $\CZ_p$ of zeroes of the $p^{\mbox{th}}$ vector field
is an invariant submanifold for~\rref{CS} as well,
so that one can restrict the flows on
{\em intersections} of invariant manifolds 
$\CS_n\cap\CZ_p$.
In particular, to study Bsq$_5$
we shall consider $\Mfive=\CS_3\cap\CZ_5$.
The restricted flows are constructed
noticing that  the coefficients of the
Laurent expansion of the equations
\begin{equation}
\dsl\dpt{}{5} h=0;\qquad
\dsl\dpt{}{5} k=0
\end{equation}
give polynomial equations for the coefficients $(h_i, k_j), i,j\ge 6$
which can be recursively solved in terms of the first
ten variables $\{ h_i, k_i, i=1,\ldots, 5\}$.\\
Accordingly, upon that substitution,
we obtain a system of polynomial vector fields $X_j$ 
defined on the phase space $\Mfive$ whose first members are displayed
in Tables~1 and 2.

\begin{table}[htb]
\protect\caption{The Vector field $X_1$}
\protect\[\boxed{
\begin{array}{rl}
\dot h_{{1}}&=k_{{1}}-2\,h_{{2}}, \\
\dot k_{{1}}&=-h_{{3}}+{h_{{1}}}^{2}-k_{{2}},
\\ \dot h_{{2}}&=-2\,h_{{3}}-{h_{{1}}}^{2}+k_{{2}},
\\ \dot k_{{2}}&=-k_{{3}}+h_{{2}}h_{{1}}-h_{{4}}-h_{{1}}k_{{1}},
\\ \dot h_{{3}}&=-2\,h_{{2}}h_{{1}}+k_{{3}}-2\,h_{{4}},
\\ \dot k_{{3}}&=-k_{{4}}-h_{{1}}k_{{2}}-h_{{5}}-k_{{1}}h_{{2}}+h_{{1}}h_{{3}},
\\ \dot h_{{4}}&=-2\,h_{{5}}-{h_{{2}}}^{2}-2\,h_{{1}}h_{{3}}+k_{{4}},
\\ \dot k_{{4}}&=-h_{{1}}k_{{3}}+h_{{2}}{h_{{1}}}^{2}
-h_{{1}}h_{{4}}+{h_{{1}}}^{2}k_{{1}}-2\,k_{{1}}h_{{3}}-k_{{1}}k_{{2}}-2\,k_{{2}}
h_{{2}},\\
\dot h_{{5}}&= 3\,k_{{5}}
-2\,k_{{1}}k_{{2}}-2\,k_{{2}}h_{{2}}-2\,h_{{2}}h_{{3}}-2\,k_{{1}}h_{{3
}}+2\,h_{{2}}{h_{{1}}}^{2}+2\,{h_{{1}}}^{2}k_{{1}}-6\,h_{{1}}h_{{4}},
\\
\dot k_{{5}}&=-3\,k_{{2}}h_{{3}}-k_{{1}}h_{{4}}-h_{{1}}{k_{{1}}}^{2}-h_{{2}}k
_{{3}}-h_{{2}}h_{{4}}+{h_{{2}}}^{2}h_{{1}}-k_{{1}}k_{{3}}
\end{array}}
\protect\]
\end{table}
\begin{table}[htb]
\protect\caption{The Vector field $X_2$}
\protect\[ \boxed{\begin{array}{rl}
\dot h_{{1}}&=-h_{{3}}+{h_{{1}}}^{2}-k_{{2}},
\\ \dot k_{{1}}&=h_{{4}}+h_{{1}}k_{{1}}
                            -h_{{2}}h_{{1}}-2\,k_{{3}},\\ \dot h_{{2}}
&=-k_{{3}}+h_{{2}}h_{{1}}-h_{{4}}-h_{{1}}k_{{1}},
\\
 \dot k_{{2}}&=-{k_{{1}}}^{2}-{h_{{2}}}^{2}+h_{{5}}-h_{{1}}k_{{2}
}-2\,k_{{4}}+2\,k_{{1}}h_{{2}},\\
\dot h_{{3}}&=-k_{{4}}-h_{{1}}k_{{2}}-h_{{5}}
-k_{{1}}h_{{2}}+h_{{1}}h_{{3}}, \\
 \dot k_{{3}}&=-3\,k_{{5}}-k_{{1}}k_{{2}}+k_{{
2}}h_{{2}}-h_{{2}}{h_{{1}}}^{2}-h_{{1}}k_{{3}}+2\,h_{{1}}h_{{4}}-{h_{{
1}}}^{2}k_{{1}}+3\,k_{{1}}h_{{3}}-h_{{2}}h_{{3}}  \ , \\
\dot h_{{4}}&=-h_{{1}}k_{{3
}}+h_{{2}}{h_{{1}}}^{2}-h_{{1}}h_{{4}}+{h_{{1}}}^{2}k_{{1}}-2\,k_{{1}}
h_{{3}}-k_{{1}}k_{{2}}-2\,k_{{2}}h_{{2}}  \ ,\\
\dot k_{{4}}&=h_{{1}}{k_{{1}}}^{2}-
{h_{{2}}}^{2}h_{{1}}+h_{{1}}h_{{5}}-2\,h_{{1}}k_{{4}}+2\,k_{{1}}h_{{4}
}-4\,k_{{1}}k_{{3}}-k_{{2}}h_{{3}}-{k_{{2}}}^{2} \ ,\\
\dot h_{{5}}&=-3\,k_{{2}}h_
{{3}}-k_{{1}}h_{{4}}-h_{{1}}{k_{{1}}}^{2}-h_{{2}}k_{{3}}-h_{{2}}h_{{4}
}+{h_{{2}}}^{2}h_{{1}}-k_{{1}}k_{{3}} \ ,\\
\dot k_{{5}}&=h_{{2}}{k_{{1}}}^{2}-{h_
{{2}}}^{3}+h_{{2}}h_{{5}}-h_{{2}}h_{{1}}k_{{2}}-2\,h_{{2}}k_{{4}}+
 k_{{1}}{h_{{2}}}^{2}-{k_{{1}}}^{3}+k_{{1}}h_{{5}}- \\
& \qquad h_{{1}}k_{{1}}k_{{2}}-2 \,k_{{1}}k_{{4}}-3\,k_{{3}}k_{{2}}
\end{array}}
\protect\]
\end{table}

\section{Lax Representation and Poisson Structures}\label{sec3}
Another consequence of Proposition~\rref{invsub} is that, (we are sticking to
the case $n=3$),
for each $j\neq 3l$ there is a non trivial $3\times 3$ matrix
$\V{j}(\lambda)$
{\em polynomially} depending on $\lambda$ such that
\begin{equation}
\left(\dpt{}{j}+\H{j}\right)
       \left[\begin{array}{c}
              1\\
              h\\
              k
              \end{array}\right]
= \V{j} (\lambda)\cdot
       \left[\begin{array}{c}
              1\\
              h\\
              k
              \end{array}\right]
\end{equation}
The commutativity of CS and the exactness property~\rref{exa}
implies the zero curvature (or Zakharov--Shabat) representation for the
restriction of CS to $\CS_3$.
\begin{equation}\label{ZS}
\dpt{}{j}\V{k} (\lambda)-\dpt{}{k}\V{j} (\lambda)+[\V{k} (\lambda),
\V{j} (\lambda)]=0
\end{equation}
The step to get a Lax problem for Bsq$_5$ on $\Mfive$ is now easy.
In fact, on the stationary manifold of $X_5=\dsl\dpt{}{5}$
we immediately obtain from~\rref{ZS}
that the {\em Lax matrix} $\V{5}(\lambda)$
satisfies the equations
\begin{equation}\label{eq:lax}
\dpt{}{k}\V{5} (\lambda)=[\V{k} (\lambda),
\V{5} (\lambda)].
\end{equation}
In other words, on the matrix $\V{5}$, all the flows of $\Mfive$
are Lax evolution equations.
Explicitly, the Lax matrix $\V{5}$ (computed using Eq.~\rref{constr})
has the following form:
\begin{equation}
\label{laxmat}
\begin{split}
\V{5}(\lambda)=&\lambda^2\ \left[
{\begin{array}{ccc}
0 & 0 & 0 \\
1 & 0 & 0 \\
0 & 1 & 0
\end{array}}
 \right] + \lambda \left[
{\begin{array}{ccc}
0 & 0 & 1 \\
{h_{2}} & {h_{1}} & 0 \\
 - {h_{3}} + {k_{2}} & {k_{1}} - {h_{2}} &  - {h_{1}}
\end{array}}
 \right] +\\
& \left[
\begin{array}{ccc}
- k_{3} & - {k_{2}} & - {k_{1}}\\
L_0^{2,1} &- {h_{2}}\,{h_{1}} + {h_{4}}&
- {h_{1}}^{2} + {h_{3}} \\
L_0^{3,1} & L_0^{3,2} & {h_{2}}\,{h_{1}} -
{h_{4}} + {k_{3}}\end{array}\right],
\end{split}
\end{equation}
where
\begin{equation}
\begin{split}
L_0^{2,1}=& - {h_{1}}\,{k_{2}} + {k_{4}} - {k_{1}}^{2} - {h
_{1}}\,{h_{3}} + {h_{5}},\\
L_0^{3,2}=&- {k_{1}}\,{h_{2}} + {h_{2}}^{2} -
{h_{5}} + 2\,{k_{4}} - {k_{1}}^{2},\\
L_0^{3,1}=&{h_{1}}\,{k_{3}} - 2\,{h_{1}}\,{h_{4}} + 3\,{k_{5}}
+ {h_{2}}\,{h_{1}}^{2}- 2\,{k_{1}}\,{k_{2}} \\ &
- 2\,{k_{1}}\,{h_{3}} +
{h_{1}}^{2}\,{k_{1}} - 2\,{k_{2}}\,{h_{2}} + {h_{2}}\,{h_{3}}.
\end{split}
\end{equation}
One can notice that the map $\{h_1,\ldots,k_5\}\mapsto \V{5}$ is invertible,
and thus the Lax
equations~\rref{eq:lax} represent faithfully Bsq$_5$.

The second members of the Lax pair for the flows $X_1$ and $X_2$
are given by
\begin{equation}
\V{1}=\left [\begin {array}{ccc} 0&1&0\\\noalign{\medskip}2\,h_{{1}}&0&1
\\\noalign{\medskip}\lambda+h_{{2}}+k_{{1}}&h_{{1}}&0\end {array}
\right ]
\quad \V{2}=\left [\begin {array}{ccc}
0&0&1\\\noalign{\medskip}\lambda+h_{{2}}+k_
{{1}}&h_{{1}}&0\\\noalign{\medskip}-h_{{3}}+2\,k_{{2}}&\lambda-h_{{2}}
+2\,k_{{1}}&-h_{{1}}\end {array}\right ]\; .
\end{equation}

\subsection{The Hamiltonian structures}
Besides their
Lax representation,
the  vector fields
of Bsq$_5$ have another important property:
they admit as well a bihamiltonian
representation, or a Poisson formulation {\em with parameter.}
We have seen that the Lax formulation
of the problem comes from the Central System. The Poisson
formulation can be gotten as follows.

As it is well known~\cite{MM,RSTS,MaMag},
on the space $\alg[[\lambda]]$
of Laurent polynomials with values in a Lie algebra
$\alg$, there is a family of mutually compatible Poisson
tensors, $\CP_l$ associated with a family of classical $R$--matrices
\begin{equation}\label{rmat}
R_l(X(\lambda))=(\lambda^l X(\lambda))_+-(\lambda^l X(\lambda))_-\quad .
\end{equation}
Some of them (including the values $l=0,1$) restrict to the
affine manifold
\begin{equation}\label{matA}
\alg_n^A:=\{X\in \alg\left[\left[\lambda\right]\right]  |
X(\lambda)=\lambda^n A+\sum_{i=0}^{n-1}\lambda^i X_i.\}
\end{equation}
where $A$ is a {\em fixed} element of $\alg$.\par
A perhaps less known fact, pointed out in~\cite{PeVa97}, is that
$\CP_0 \mbox{ and } \CP_1$  admit further reductions,
leading to generalized Mumford systems.
Actually we want to show that
the Hamiltonian structures of Bsq$_5$, gotten via the reduction
of the central system discussed in Section~\ref{sec2},
are a Marsden--Ratiu bihamiltonian
reduction of the pair
$ \CP_0 \mbox{ and } \CP_1$. This choice of Poisson pair is by no
means accidental.
In fact a bihamiltonian equation on $\alg_n^A$
\begin{equation}
\dot X= (\CP_1-\lambda\CP_0) \nabla F
\end{equation}
imply that the polynomial $X(\lambda)$ evolves according 
to a Lax equation

We recall that the MR reduction theorem~\cite{MR} considers a Poisson
manifold $(\CM, P)$, a submanifold $\CS\hookrightarrow\CM$
and a distribution $D\subset T\CM_{\vert_\CS}$ such that
$E=D\cap T\CS$ is a regular foliation with a good quotient
$\CN= \CS/E$.
It states that the Poisson tensor $P$ is reducible to $\CN$ if
the following hold:
\begin{enumerate}
\item The functions on $M$ invariant along $D$ form a
Poisson subalgebra of $C^\infty(\CM)$.
\item
$P(D^0)\subset T\CS + D$, $D^0$ being the annihilator of $D$ in $T^*\CM$.
\end{enumerate}
The reduced Poisson tensors can be computed
according to the following scheme:
\begin{itemize}
\item We choose  a covector $v_n^\CN\in T^*_n\CN$.
\item We choose a point $s\in\CS$ on the fiber over $n$,
and we lift $v_n^\CN$ to a covector $v_s^\CM\in T^*_s\CM$,
that is an extension of $\pi^*v_n^\CN$ lying in the
annihilator $D^0$ of the distribution $D$.
\item Next we construct the vector field $(P^\CM)_sv_s^\CM$
associated with the lifted covector $v_s^\CM$ through the Poisson
tensor of $\CM$ at the point $s$. The MR reduction theorem 
insures that
the vector $(P^\CM)_sv_s^\CM$ is projectable onto $\CN$.
The projection  does not
depend either on the choice of the particular extension $v^\CM_s$
or on the point $s$ on the fiber and defines
$(P^\CN)_nv_n^\CN$.
\end{itemize}

The {\em \bih\ MR reduction theorem}~\cite{CMP} 
considers a manifold $( \CM, P_0, P_1 )$ endowed with a pair 
of compatible Poisson structures, i.e. a {\em \varb.} 
It is a consequence of the MR theorem
stemming from the observation that a \bih\ \var\ admits a
kind of {\em double} foliation. This suggests to
choose $\CS$ to be a symplectic leaf of $P_0$ and
$D=P_1(\mbox{Ker}P_0)$. Then, provided that $\CS$ is
chosen so that the regularity assumption of the MR theorem
is satisfied, points 1) and 2) above are
consequences of the compatibility of the Poisson pair $(P_0,
P_1)$.
Moreover, it states the reducibility of the
whole Poisson pencil $P_\lambda=P_1-\lambda P_0$, and provides $\CN$ with a
bihamiltonian structure which can be computed following the procedure 
sketched above.

\bigskip
In our Bsq$_5$ problem we consider the reduction of the Poisson pencil
$\CP_\lambda=\CP_1-\lambda\CP_0$, associated with the
$R$--matrices~\rref{rmat}, defined on the space 
$3\times 3$ traceless matrices of the form
$X(\lambda)=\lambda^2 A+\lambda x_1+x_0$,
with
\[
A=\left[
\begin{array}{ccc}
0&0&0\\
1&0&0\\
0&1&0
\end{array}\right].
\]
Making use of the Killing form on $\fraksl(3)$, and of the residue in
$\lambda$, we can
parameterize tangent and cotangent vectors as matrix--valued (Laurent)
polynomials as well:
\[
\dot X=\lambda\dot x_1+\dot x_0;\qquad
W=\frac{w_0}{\lambda}+\frac{w_1}{\lambda^2}.
\]
The Poisson tensors  $\CP_0$ and $\CP_1$ have the explicit form~\cite{MaMag}:
\begin{equation}\label{p0mat}
\dot X = \CP_0(W)  \Leftrightarrow \left( \begin{array}{c} \dot x_0\\\dot
x_1\end{array} \right) = \left(
\begin{array}{cc} \left[x_1, \cdot\right] & \left[A, \cdot\right] \\
\left[A, \cdot\right] &0
\end{array}\right)
\left( \begin{array}{c} w_0 \\ w_1 \end{array} \right)
=\left(
\begin{array}{c} \left[x_1,w_0\right] + \left[A,w_1\right] \\
\left[A,w_0\right] \end{array} \right)
\end{equation}
and
\begin{equation}\label{p1mat}
\dot X = \CP_1(W)  \Leftrightarrow \left( \begin{array}{c} \dot x_0\\\dot
    x_1\end{array} \right) = \left(\begin{array}{cc} -\left[x_0,
\cdot\right] &
    0 \\ 0 &\left[A, \cdot\right]\end{array}\right)
\left( \begin{array}{c} w_0\\ w_1
   \end{array} \right)=\left( \begin{array}{c} \left[w_0,x_0\right]\\
{}\left[A,w_1\right]  \end{array} \right).
\end{equation}
A  long but straightforward computation shows that the Lax matrix $\V{5}$ fits
into the  scheme herewith outlined, so that via the \bih\ MR reduction process
the space $\Mfive$ is endowed
with a Poisson pencil $P_\lambda=P_1-\lambda P_0$;
in the coordinates $(h_1,k_1, h_2, k_2, \dots, h_5, k_5)$  $P_0$ and $P_1$
have the form displayed in Tables 3 and 4.
\begin{table}
\protect\caption{The reduced tensor $P_0$}
\protect\[\label{p0}
\boxed{
{\begin{array}{rrrrcccccc}
0 & 0 & 0 & 0 & 0 & 0 & 0 & 0 & -1 & 0 \\
& 0 & 0 & 0 & 0 & 1 & 0 & 0 & 0 & 0 \\
& & 0 & 0 & 0 & 1 & 1 & 0 & 0 & 0 \\
& & & 0 & 1 & 0 & 0 & 0 & 0 & 0 \\
& & & & 0 & 0 & 0 &  - {h_{1}} &  - 2\,{h_{1}} &  - {k_{1}} - {h_{2}} \\
&&&&& 0 &  - {h_{1}} &  - 2\,{k_{1}} &  - {k_{1}} - {h_{2}} &  - 2\,{k_{2}} \\
&&*&&&& 0 & 0 &  - 2\,{h_{2}} &  - {k_{2}}
 \\
&&&&&&& 0 &  - {k_{2}} & 0 \\
&&&&&&&& 0 & 0 \\
&&&&&&&&& 0
\end{array}}}
\protect\]
\end{table}
\begin{table}[ht]
\protect\caption{The reduced tensor $P_1$.
The entries $P^{l,k}$ are the polynomials in the coordinates
$\{h_1,\ldots,k_5\}$ written below}
\protect\[
\frac{1}{3}\boxed{
\begin {array}{cccccccccc} 0&0&-3&0&0&0&-3h_{1}&
2h_{2}-k_{1}&k_{1}-2h_{2}&P^{1,10}
\\&0&0&0&0&3k_{1}&0&h_{3}-{h_{1}}^{2}+k_{2}
&2h_{3}-2{h_{1}}^{2}-k_{2}&P^{2,10}\\&
&0&0&6h_{1}&3k_{1}&3h_{2}&2k_{2}+2h_{3}+{h_{1}}^{2}&
k_{2}+4h_{3}+2{h_{1}}^{2}&P^{3,10}\\&&
&0&3k_{1}&3k_{2}&3k_{2}&P^{4,8}&P^{4,9}&P^{4,10}
\\&&&&0&0&-3{h_{1}}^{2}-3h_{3}&P^{5,8}&
k_{3}-2h_{2}h_{1}-2h_{4}&P^{5,10}\\&&&
&&0&-3h_{1}k_{1}&P^{6,8}&P^{6,9}&P^{6,10}
\\&&*&&&&0&P^{7,8}&P^{7,9}&P^{7,10}
\\&&&&&&&0&P^{8,9}&P^{8,10}
\\&&&&&&&&0&P^{9,10}
\\&&&&&&&&&0\end {array}}
\protect\]
\vspace{.5truecm}
\protect\[\boxed{\begin{array}{l}
P^{1,10}=h_{3}-2k_{2}-{h_{1}}^{2},\qquad
P^{2,10}=-h_{1}k_{1}+h_{2}h_{1}-h_{4}+2k_{3},\\
P^{3,10}=k_{3}+h_{4}-h_{2}h_{1}+h_{1}k_{1},\qquad
P^{4,8}=k_{3}+h_{4}-h_{2}h_{1}+h_{1}k_{1},\\
P^{4,9}=2k_{3}+2h_{4}-2h_{2}h_{1}+2h_{1}k_{1},
\quad P^{4,10}=-2k_{1}h_{2}+{k_{1}}^{2}+h_{1}
k_{2}-h_{5}+{h_{2}}^{2}+2k_{4},\\
P^{5,8}=-3h_{1}k_{1}+2h_{2}h_{1}+2h_{4}-k_{3},\qquad
P^{5,10}=-h_{1}h_{3}-5h_{1}k_{2}-2k_{1}h_{2}-3{k_{1}}^{2}+h_{5}+k_{4}\\
P^{6,8}=-6k_{1}^{2}+
k_{1}h_{2}-2h_{1}k_{2}+k_{4}-h_{1}h_{3}+h_{5},\\
P^{6,9}=-3{k_{1}}^{2}-k_{1}h_{2}-4h_{1}k_{2}
+2k_{4}-2h_{1}h_{3}+2h_{5},\\
P^{6,10}=-3k_{1}h_{3}-8k_{1}k_{2}-4k_{2}
h_{2}+h_{1}k_{3}+h_{2}h_{3}-2h_{1}h_{4}+h_{2}{h_{1}}^{2}+3k_{5}+
{h_{1}}^{2}k_{1},\\
P^{7,8}={h_{2}}^{2}+2h_{1}h_{3}-k_{4}+2h_{5}, \qquad
P^{7,9}=-{h_{2}}^{2}-3h_{1}k_{2}+4h_{1}h_{3}-3{k_{1}}^{2}+4k_{4}
+h_{5},\\
P^{7,10}=h_{1}k_{3}+h_{1}h_{4}-h_{2}{h_{1}}^{2}-{h_{1}}^{2}k_{1}-4k_{2}h_{2}-2k_
{1}k_{2}+2k_{1}h_{3},\\
P^{8,9}=-4k_{2}h_{2}-4h_{1}h_{4}
+2k_{1}h_{3}-3k_{1}k_{2}+3k_{5}+2h_{1}k_{3}-2
h_{2}h_{3},\\
P^{8,10}=-2{k_{2}}^{2}
-h_{1}h_{5}+2h_{1}k_{4}-2k_{2}h_{3}-3k_{1}h_{4}-2
h_{1}{k_{1}}^{2}-h_{2}k_{3}-h_{2}h_{4}+2{h_{2}}^{2}h
_{1}+3k_{1}k_{3},\\
P^{9,10}=-3k_{2}h_{3}-k_{1}h_{4}-h_{1}{k_{1}}^{2}
-h_{2}k_{3}-h_{2}h_{4}+{h_{2}}^{2}h_{1}-k_{1}k_{3}
\end{array}}
\protect\]
\end{table}

We remark that, in the case of the $t_4$--stationary Boussinesq hierarchy,
this procedure yields a pair of Poisson structures which coincide with the
ones found in~\cite{FH}.

The hamiltonian functions for the flows can in principle
be computed looking for the Casimir function of
the Poisson pencil $P_1-\lambda P_0$, according the Gel'fand--Zakharevich
theorem~\cite{GZ93}. However, since we are dealing with a Lax
problem with spectral parameter, they are
provided as well by the
coefficients of the spectral curve associated with the problem, i.e:
\begin{equation}\label{spec_cur}
0=\mbox{det}(\mu-\V{5})=\mu^3-\mu(H_1
\lambda+H_2)-(\lambda^5+H_3\lambda^3+H_4\lambda^2+H_5\lambda+H_6)
\end{equation}
The explicit expressions of the Hamiltonian functions are:
\begin{small}
\begin{equation}\label{GZham}
\begin{split}
H_1=&3k_{{5}}-3k_{{1}}k_{{2}}-3k_{{2}}h_{{2}}, \quad
H_3=-3h_{{1}}k_{{2}}+3k_{{4}}-3{k_{{1}}}^{2},\\
H_2=&-3k_{{5}}k_{{1}}-h_{{5}}k_{{2}}
+h_{{3}}{k_{{1}}}^{2}-2h_{{2}}h_{{1}}h_{{4}}-
h_{{4}}k_{{3}}+h_{{3}}{h_{{2}}}^{2}-h_{{3}}h_{{5}}-\\ & h_{{1}}k_{{3}}k_{{1
}}+h_{{2}}h_{{1}}k_{{3}}+h_{{1}}h_{{3}}k_{{2}}+ h_{{1}}{k_{{2}}}^{2}+3
{k_{{1}}}^{2}k_{{2}}-k_{{4}}k_{{2}}+2k_{{2}}h_{{2}}k_{{1}}+\\ &{k_{{3}
}}^{2}+{h_{{4}}}^{2}-2{h_{{1}}}^{2}k_{{4}}+2h_{{3}}k_{{4}}-2h_{{
3}}k_{{1}}h_{{2}}+2h_{{1}}h_{{4}}k_{{1}}+{h_{{1}}}^{2}h_{{5}}.\\
H_4=&-2k_{{1}}k_{{4}}-2k_{{2}}h_{{4}}+k
_{{3}}h_{{3}}-3h_{{1}}k_{{5}}-2h_{{2}}h_{{5}}+3h_{{1}}k_{{1}}k_{
{2}}+\\ & 4h_{{2}}h_{{1}}k_{{2}}+{h_{{2}}}^{3}+{k_{{1}}}^{3}-h_{{1}}h_{{2
}}h_{{3}}+h_{{2}}k_{{4}}+h_{{4}}h_{{3}}-{h_{{1}}}^{2}k_{{3}}+\\ & k_{{1}}h_
{{5}}+ h_{{1}}h_{{3}}k_{{1}}-k_{{1}}{h_{{2}}}^{2}+k_{{3}}k_{{2}}-h_{{2}
}{h_{{1}}}^{3}-{h_{{1}}}^{3}k_{{1}}+2{h_{{1}}}^{2}h_{{4}}.
\end{split}
\end{equation}
\end{small}
The functions $H_5$ and $H_6$ are written in Table~5.

\begin{equation*}
\boxed{
\begin{split}
&\qquad\qquad\qquad\qquad\text{Table 5: The Hamiltonians $H_5$ and $H_6$} \\ &
\\ H_5=&h_{{1}}{k_{{1}}}^{2}k_{{2}}+h_{{4}}h_{{3}}k_{{1}}+3\,k_{{1}}k_
{{2}}h_{{4}}+2\,h_{{2}}{h_{{1}}}^{3}k_{{1}}+h_{{1}}h_{{5}}h_{{3}}-3\,h
_{{1}}k_{{2}}k_{{4}} +3\,h_{{1}}k_{{1}}k_{{5}}+h_{{1}}h_{{4}}k_{{3}}\\&-2
\,h_{{2}}k_{{1}}k_{{4}}+k_{{3}}k_{{2}}h_{{2}}-3\,h_{{2}}{h_{{1}}}^{2}h
_{{4}}+k_{{2}}h_{{2}}h_{{4}}-2\,h_{{1}}k_{{4}}h_{{3}}-2\,{h_{{2}}}^{2}
h_{{1}}k_{{2}}-3\,k_{{1}}{h_{{1}}}^{2}h_{{4}}\\&+3\,h_{{2}}h_{{1}}k_{{5}}
+h_{{3}}k_{{1}}k_{{3}}+{h_{{2}}}^{2}h_{{5}} +h_{{5}}k_{{4}}-h_{{5}}{k_{
{1}}}^{2}+{h_{{3}}}^{2}k_{{2}} +{h_{{2}}}^{2}{h_{{1}}}^{3}-h_{{1}}{k_{{
3}}}^{2}\\&-{k_{{2}}}^{2}h_{{3}}+{h_{{1}}}^{3}{k_{{1}}}^{2}+ h_{{2}}{k_{{1
}}}^{3}+{h_{{2}}}^{2}k_{{4}}-k_{{1}}{h_{{2}}}^{3}-3\,h_{{4}}k_{{5}}-4
\,{k_{{1}}}^{2}k_{{4}}+2\,{h_{{1}}}^{2}{k_{{2}}}^{2}\\&+2\,h_{{1}}{h_{{4}
}}^{2}+ k_{{1}}h_{{2}}h_{{5}}-h_{{2}}h_{{3}}k_{{3}}-h_{{2}}h_{{4}}h_{{3
}}+2\,{k_{{4}}}^{2}+2\,{k_{{1}}}^{4}-{h_{{5}}}^{2}-5\,h_{{2}}h_{{1}}k_
{{1}}k_{{2}}\\
H_6=&-{k_{{1}}}^{2}k_{{2}}h_{{4}}-h_{{1}}{k_{{1}}}^{3}k_{{2}}-3\,h_{{2}}{h_
{{1}}}^{2}{k_{{2}}}^{2}-3\,k_{{5}}k_{{2}}h_{{3}}-{k_{{1}}}^{2}h_{{2}}{
h_{{1}}}^{3}+2\,k_{{1}}{k_{{2}}}^{2}h_{{3}}+{h_{{1}}}^{2}{k_{{1}}}^{2}
k_{{3}}\\ & -k_{{2}}h_{{4}}k_{{4}}+h_{{1}}{k_{{2}}}^{2}h_{{4}}-h_{{5}}k_{{1
}}k_{{4}}+2\,{k_{{2}}}^{2}h_{{2}}h_{{3}}-2\,k_{{4}}{h_{{1}}}^{2}k_{{3}
}+{k_{{1}}}^{2}h_{{2}}h_{{5}}+{h_{{1}}}^{4}k_{{2}}h_{{2}}\\ & -{k_{{1}}}^{2
}k_{{2}}k_{{3}} -2\,h_{{1}}k_{{1}}{h_{{4}}}^{2}+2\,k_{{1}}{h_{{3}}}^{2}
k_{{2}}+{k_{{1}}}^{2}{h_{{1}}}^{2}h_{{4}}-{k_{{1}}}^{2}h_{{3}}k_{{3}}+
k_{{4}}k_{{3}}k_{{2}}+3\,h_{{4}}k_{{1}}k_{{5}}\\ & +2\,k_{{4}}h_{{3}}k_{{3}
}-2\,h_{{4}}h_{{3}}{k_{{1}}}^{2}-3\,h_{{2}}h_{{1}}k_{{1}}k_{{5}}-h_{{1
}}k_{{1}}k_{{2}}h_{{5}}+h_{{2}}h_{{1}}k_{{2}}h_{{5}}+2\,h_{{1}}k_{{1}}
k_{{2}}k_{{4}}\\ & -2\,k_{{1}}k_{{2}}h_{{2}}h_{{4}}+3\,{h_{{2}}}^{2}h_{{1}}
k_{{1}}k_{{2}} +h_{{2}}h_{{1}}h_{{3}}{k_{{1}}}^{2}-2\,h_{{2}}h_{{1}}h_{
{4}}k_{{3}}-h_{{2}}{h_{{1}}}^{2}k_{{2}}h_{{3}}+3\,h_{{2}}{h_{{1}}}^{2}
k_{{1}}h_{{4}}\\ & -2\,h_{{1}}k_{{2}}h_{{3}}k_{{3}}+2\,h_{{1}}h_{{3}}k_{{1}
}k_{{4}}+3\,h_{{1}}k_{{2}}h_{{4}}h_{{3}} -3\,{h_{{1}}}^{2}k_{{2}}h_{{3}
}k_{{1}}-k_{{1}}{h_{{2}}}^{2}k_{{4}}+3\,{h_{{1}}}^{2}k_{{2}}k_{{5}}\\ & -k_
{{2}}h_{{4}}h_{{5}}+k_{{3}}k_{{2}}h_{{5}}+h_{{2}}h_{{1}}{k_{{3}}}^{2}+
h_{{5}}{h_{{1}}}^{2}k_{{3}}-h_{{5}}h_{{3}}k_{{3}}+{h_{{2}}}^{2}h_{{3}}
k_{{3}}-h_{{2}}{h_{{3}}}^{2}k_{{2}}\\ & -k_{{1}}{h_{{2}}}^{2}h_{{5}}-k_{{1}
}{h_{{2}}}^{2}{h_{{1}}}^{3}-h_{{1}}{k_{{2}}}^{2}k_{{3}}+{h_{{1}}}^{4}k
_{{2}}k_{{1}} -2\,{h_{{1}}}^{3}k_{{2}}h_{{4}}+{k_{{1}}}^{2}h_{{2}}k_{{4
}}-h_{{1}}h_{{3}}{k_{{1}}}^{3}\\ & +{h_{{1}}}^{3}k_{{2}}k_{{3}}-2\,{h_{{1}}
}^{2}{k_{{2}}}^{2}k_{{1}}+3\,{k_{{1}}}^{3}k_{{4}}+{h_{{4}}}^{2}k_{{3}}
-h_{{4}}{k_{{3}}}^{2}+k_{{1}}{h_{{5}}}^{2}-2\,k_{{1}}{k_{{4}}}^{2}+{k_
{{1}}}^{3}{h_{{2}}}^{2}\\ & -{k_{{1}}}^{4}h_{{2}}+h_{{2}}h_{{1}}k_{{2}}k_{{
4}}-k_{{1}}h_{{2}}h_{{3}}k_{{3}}-h_{{1}}k_{{1}}h_{{5}}h_{{3}}+h_{{1}}h
_{{4}}k_{{1}}k_{{3}}+k_{{1}}h_{{2}}h_{{4}}h_{{3}}-{k_{{1}}}^{5}
\end{split}}
\end{equation*}
A direct computation shows  that
\begin{prop}\label{prop:GZ}
The coefficients $H_j$ of the characteristic polynomial of the Lax matrix
$\V{5}(\lambda)$,
the vector fields $(X_1,X_2,X_4,X_7)$,
and the Poisson tensors $P_0$ and $P_1$ of Tables~3 and~4 fill 
the bihamiltonian sequences of Figure~1.
In particular, the hamiltonian functions
$H_j$ are mutually in involution w.r.t. the whole Poisson pencil.
\end{prop}
\begin{figure}[hb]
\begin{center}
  \caption{The Bihamiltonian Sequences}
\end{center}
\setlength{\unitlength}{2500sp}
\begin{center}
\thicklines
\begin{picture}(8475,3345)(1426,-2761)
\put(2401,-1861){\vector(-1,-1){675}}
\put(2401,239){\vector(-1,-1){675}}
\put(2701,239){\vector( 1,-1){675}}
\put(4501,239){\vector(-1,-1){675}}
\put(4801,239){\vector( 1,-1){675}}
\put(2701,-1861){\vector( 1,-1){675}}
\put(4501,-1861){\vector(-1,-1){675}}
\put(4801,-1861){\vector( 1,-1){675}}
\put(6601,-1861){\vector(-1,-1){675}}
\put(6901,-1861){\vector( 1,-1){675}}
\put(9001,-1861){\vector( 1,-1){675}}
\put(8701,-1861){\vector(-1,-1){675}}
\put(1426,-661){\makebox(0,0)[lb]{0}}
\put(1501,-2761){\makebox(0,0)[lb]{$0$}}
\put(3500,-661){\makebox(0,0)[lb]{$X_2$}}
\put(3500,-2761){\makebox(0,0)[lb]{$X_1$}}
\put(5600,-2761){\makebox(0,0)[lb]{$X_4$}}
\put(7700,-2761){\makebox(0,0)[lb]{$X_7$}}
\put(9875,-2761){\makebox(0,0)[lb]{$0$}}
\put(5675,-661){\makebox(0,0)[lb]{$0$}}
\put(2401,389){\makebox(0,0)[lb]{$dH_1$}}
\put(4501,389){\makebox(0,0)[lb]{$dH_2$}}
\put(2401,-1786){\makebox(0,0)[lb]{$dH_3$}}
\put(4501,-1786){\makebox(0,0)[lb]{$dH_4$}}
\put(6601,-1786){\makebox(0,0)[lb]{$dH_5$}}
\put(8701,-1786){\makebox(0,0)[lb]{$dH_6$}}
\put(1801, 14){\makebox(0,0)[lb]{$P_0$}}
\put(3076, 14){\makebox(0,0)[lb]{$P_1$}}
\put(1876,-2086){\makebox(0,0)[lb]{$P_0$}}
\put(3076,-2086){\makebox(0,0)[lb]{$P_1$}}
\end{picture}
\end{center}
\end{figure}
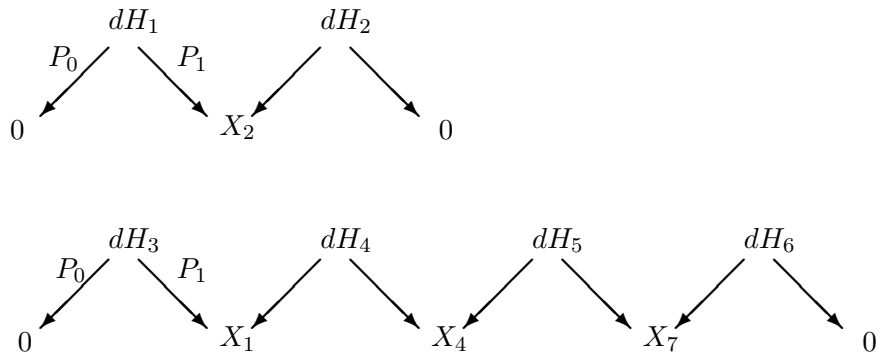
\section{PN manifolds and the Sklyanin procedure}
\label{sec4}
In this Section
we show how the bihamiltonian structure of the
Bsq$_5$ hierarchy
can be exploited in order to get a set of variables which satisfy
Sklyanin's separation equations.
Actually,  this kind of variables arise as a set
of coordinates which put in ``canonical'' form a Poisson--Nijenhuis (PN)
structure obtained by a further Marsden--Ratiu reduction of the degenerate
Poisson pencil $P_\lambda$.

\subsection{The reduction of the Poisson pencil}

Let $M$ be a bihamiltonian manifold with degenerate  Poisson tensors $(P_0,
P_1)$ and let $X_j$ be the vector fields of a bihamiltonian  sequence on $M$.
A possible way to analyze the  integrability of  $X_j$ is to eliminate
the Casimir functions of one Poisson tensor, say $P_0$, by
fixing the values of its Casimirs.
Of course, $P_0$ can be restricted to any of its symplectic
leaf $S_0$ and the $X_j$, being tangent
to $S_0$,  can be restricted as well to  vector fields
$\widehat{X}_j$. These are still Hamiltonian on
$S_0$, with Hamiltonians $\widehat{H}_j$ given 
by the {\em restrictions} of the original Hamiltonians to $S_0$.
However, in general, the bihamiltonian 
formulation of the restricted problem is lost on $S_0$.
\par
Nonetheless, in our situation a specific feature of the pencil allows
us to  reduce the bihamiltonian structure
to any generic leaf $S_0$ of $P_0$.
Indeed, the Poisson pencil $P_\lambda$ induces on $S_0$
a Poisson--Nijhenuis (PN) structure~\cite{MM,FoFu}. 
With respect to such a PN structure, 
the  fields $\widehat{X}_j$ admit a formulation, to be introduced in Section
\ref{sec5}, powerful enough to allow us to find a set of separation variables.

\begin{lem}\label{ld:lem}
Let us consider the vector fields
\begin{equation}
Z_1=\frac{\del}{\del k_5},\quad Z_2=\frac{\partial}{\partial
  k_4}+2\frac{\partial}{\partial h_5},
\end{equation}
and let $\widehat{\CD}$ be the distribution generated by $Z_1\, \mbox{and}\,
Z_2$.
Then it holds that:
\begin{equation}
L_{\phi_1 Z_1+ \phi_2Z_2}(P_1-\lambda P_0)=Z_1\wedge W_1+Z_2\wedge W_2;
\end{equation}
where $L_X P$ is the Lie derivative of $P$ along $X$, $\phi_1$ and $\phi_2$ 
are generic smooth functions and the
$W_i$'s are suitable vector fields (which depend on $\phi_1$ and $\phi_2$)
\end{lem}
This condition ensures that the space of functions annihilated by
$\widehat{\CD}$ is actually a {\em Poisson subalgebra} of $C^\infty(\Mfive)$.
Moreover $\widehat{\CD}$ is {\em generically transversal} to 
the image of $P_0$. Hence we can apply
the MR theorem to state
\begin{prop}\label{prpo4}
The pencil $P_\lambda$ is
{\em reducible} to a pencil $\widehat{P}_\lambda$ on $S_0$.
Since $\widehat{P}_0$ is invertible on $S_0$, that is,
$\omega=\widehat{P}_0^{-1}$ is a {\em symplectic} two--form, 
we can conclude that $S_0$ admits the
structure of a PN manifold, with respect to the pair
$(\widehat{P}_0,\, N=\widehat{P}_1\widehat{P}_0^{-1}).$
\end{prop}
It is convenient to adapt the coordinates on
$\CM$ to the distribution $\widehat{\CD}$
as $\{ h_1,\ldots,h_4, g_1=2 k_4-h_5, v_1=H_1, v_2=H_3\}.$  
In this way,  the reduced Poisson tensors
 can be obtained from $P_0$ and $P_1$ (written in the 
new coordinate system) simply deleting the last two rows and columns.  
They are represented  by:
\begin{equation}   \label{eq:P0}
\widehat{P}_0=\left[
\begin{array} {cccccccc}
0& 0& 0& 0& 0&  0& 0& -1 \\
0& 0& 0& 0& 0& -1& 0& 0 \\
0& 0& 0& 0& 0& -1& -1& 0 \\
0& 0& 0& 0& -1& 0& 0 &0  \\
0& 0& 0& 1& 0& 0& 0& 0 \\
0& 1& 1& 0& 0& 0& h_1& 3k_1-h_2 \\
0& 0& 1& 0& 0& -h_1& 0& -2h_2 \\
1& 0& 0& 0& 0&  -3k_1+h_2 & 2h_2 &0
\end{array}
\right] \ ,
\end{equation}
\begin{equation}   \label{eq:P1}
\widehat{P}_1=\left[
\begin{array} {cccccccc}
0& 0& 1& 0& 0&  0& h_1& k_1-2h_2 \\
& 0& 0& 0& 0&-k_1& 0& -k_2 \\
& & 0& 0& -2h_1& -k_1& -h_2& -k_2 \\
&&& 0& -k_1& -k_2&-k_2 &0  \\
&&&& 0& 0&h_3+h_1^2& -2h_4+k_3+2h_1 k_1-2h_2 h_1 \\
&&&*&& 0& h_1 k_1& 3k_1^2-h_2 k_1\\
&&&&&& 0&g_1-h_1 k_2-k_1^2-h_2^2  \\
&&&&&&&0
\end{array} 
\right] \ .
\end{equation}
The adjoint of the Nijenhuis tensor is
\begin{equation} \label{eq:N}
N^*=
\left[
\begin {array}{cccccccc} -k_{{1}}&k_{{2}}&k_{{2}}&0&N^{{1,5}}&0&N^{{1,
7}}&3\,k_{{1}}k_{{2}}-k_{{2}}h_{{2}}\\ 0&k_{{1}}&4\,k
_{{1}}&0&-{h_{{1}}}^{2}+h_{{3}}&0&N^{{2,7}}&N^{{2,8}}
\\ -h_{{1}}&0&-h_{{2}}&k_{{2}}&-h_{{3}}-{h_{{1}}}^{2}
&0&N^{{3,7}}&N^{{3,8}}\\ 0&0&2\,h_{{1}}&k_{{1}}&0&0&h
_{{3}}+{h_{{1}}}^{2}&N^{{4,8}}\\ 0&0&0&0&k_{{1}}&k_{{
2}}&k_{{2}}&0\\ 0&0&0&0&0&k_{{1}}&0&k_{{2}}
\\ 1&0&0&0&2\,h_{{1}}&0&h_{{2}}&0\\ 0
&0&-1&0&0&0&-h_{{1}}&2\,h_{{2}}-k_{{1}}\end {array}\right ] \ ,
\end{equation}
where
\begin{equation*}
\begin{split}
N^{{1,5}}&=-2\,h_{{2}}h_{{1}}+2\,h_{{4}}-k_{{3}}-2\,h_{
{1}}k_{{1}},\quad N^{{1,7}}=-{h_{{2}}}^{2}-g_{{1}}+h_{{1}}k_{{2}}+{k_{{1}}}^
{2},\\
N^{{2,8}}&=-6\,h_{{2}}k_{{1}}-{h_{{2}}}^{2}+7\,{k_{{1}}}^{2}-g_{{1}
},\quad N^{{3,8}}=3\,{h_{{2}}}^{2}-2\,h_{{2}}k_{{1}}+g_{{1}}-{k_{{1}}}^{2},\\
N^{{4,8}}&=-2\,h_{{4}}+k_{{3}}+2\,h_{{1}}k_{{1}}-2\,h_{{2}}h_{{1}},\quad N^{{2
,7}}=4\,h_{{1}}k_{{1}},\quad N^{{3,7}}=-2\,h_{{2}}h_{{1}}\\
\end{split}
\end{equation*}

The separation of variables (SoV) is an outcome of the spectral analysis
of $N^*$, and relies on the introduction of some special set of
coordinates. This will be the subject of the next subsection.

\subsection{ The Darboux--Nijenhuis and the Hankel--Fr\"obenius coordinates}
Let $(M, P_0, N=P_1P_0^{-1})$ be a PN manifold of dimension $2n$.
In \cite{MaMar}, it was proved that  if $N$ has $n$
functionally independent eigenvalues
$(\lambda_1,\ldots ,\lambda_n)$ (otherwise stated, when $N$ 
is {\em maximal}) one can introduce a set of
coordinates
 ($ \mbox{\boldmath $\lambda$}:=(\lambda_1,\ldots,\lambda_n);
\mbox{\boldmath  $\mu$}:=(\mu_1,\ldots,\mu_n)$)  such that
$P_0$  and $N$ take the  form
\begin{equation} \label{eq:PND} P_0= \left[
\begin{array}{cc}
  0& \boldsymbol{I}\\
 - \boldsymbol{I}&0
\end{array} \right] \ ,
\qquad N=\left[
\begin{array}{cc}
\boldsymbol{\Lambda}&0 \\
 0&\boldsymbol{\Lambda}
\end{array} \right] \ ,
\end{equation} where $ \boldsymbol{I}$
is the $n \times n$ identity matrix and
$\boldsymbol{\Lambda}
:=diag(\lambda_1, \ldots, \lambda_n)$. It was also remarked that
while the first $n$ coordinates are the eigenvalues of $N$
the remaining ones can in general be constructed by quadratures.
Since these coordinates are canonical
w.r.t. $P_0$ and diagonalize the Nijenhuis tensor
they will be referred to as
{\it Darboux-Nijenhuis} (DN) coordinates.

Another set of  coordinates,  denoted as
($ \mbox{\boldmath $f$}:=(f_1,\ldots,f_n)$;
$\mbox{\boldmath $c$}:=(c_1,\ldots,c_n)$)
can be fruitfully introduced.
The idea is the following:
(minus) the coefficients $c_j$ of the minimal polynomial of the Nijenhuis
tensor
\begin{equation} \label{eq:Nmp}
\left(\mbox{Det}(\lambda-N)\right)^{\frac{1}{2}}=\lambda^n
-\sum_{i=1}^{n}c_{i}\lambda^{n-i}
\end{equation}
satisfy
the following remarkable recursion property:
\begin{equation}\label{c:rec}
\begin{array}{rcl}
N^* dc_1&= &dc_2+c_1 dc_1\\
N^* dc_2&= &dc_3+c_2 dc_1\\
&\vdots&\\
N^*dc_n&=&\quad c_n dc_1.
\end{array}
\end{equation}
In complete analogy, we seek for a set of complementary variables
$\{f_1, \ldots, f_n\}$
such that\begin{enumerate}
\item their differentials generate an $n$--dimensional distribution in
 $T^*S_0$, {\em complementary} to the one generated by the $dc_i$'s;
\item
they satisfy the same recursion relation:
\begin{equation}\label{f:rec}
\begin{array}{rcl}
N^* df_1&=& df_2+c_1 df_1\\
N^* df_2&=& df_3+c_2 df_1\\
&\vdots&\\
N^*df_n&=&\quad c_n df_1
\end{array}.
\end{equation}
\end{enumerate}
A characteristic feature of the coordinates $(\boldsymbol{f}; \boldsymbol{c})$
is that $P_0$ and
$N^*$ take the matrix  form
\begin{equation} \label{eq:PNF}
 P_0= \left[
\begin{array}{cc}
  0& \boldsymbol{H}\\
 - \boldsymbol{H}^T&0
\end{array} \right] \ ,
\qquad N^*=\left[
\begin{array}{cc}
\boldsymbol{F}&0 \\
 0&\boldsymbol{F}
\end{array} \right] \ ,
\end{equation}
where
\begin{small}
\begin{equation} \label{eq:ABCD}
\boldsymbol{H}= \left[
\begin{array}{cccc}
  0&0&\cdots &1\\
 0&\cdots&1&-c_1 \\
1&-c_1&\cdots&-c_{
n-1}
\end{array} \right] \ , \qquad
\boldsymbol{F}=\left[
\begin{array}{cccc}
c_1&c_2&\cdots&c_n \\
 1&0&\cdots&0 \\
0&\cdots&1&0
\end{array} \right] \ ,
\end{equation}
\end{small}
whence the denomination of {\em Hankel--Fr\"obenius} (HF) coordinates.

It is not difficult to show that
the relation between the $(\mbox{\boldmath  $f$};\mbox{\boldmath $ c$)}$
coordinates and the DN coordinates $(\boldsymbol{\lambda};
\boldsymbol{\mu})$ is given by the  system
\begin{equation}\label{mula:f}
\begin{split}
&\lambda^n
-\sum_{i=1}^{n}c_{i}\lambda^{n-i}=\prod_{i=1}^n
(\lambda-\lambda_i) \ \\
&\mu_k=\sum_{i=1}^{n}f_{i}\lambda_k^{n-i} \quad (k=1,\cdots,n).
\end{split}
\end{equation}
The advantage of using the Hankel--Fr\"obenius coordinate system
is that, in the cases at hand, they can be computed 
in a {\em  algebraic fashion} from the recurrence~\rref{f:rec}.
\par
As a final remark, we note that, owing to the last equations~\rref{c:rec} 
and~\rref{f:rec}, the Hamiltonian fields $X=P_0 dc_1$ and $Y=P_0 df_1$ are
Pfaffian quasi--bihamiltonian vector fields in the terminology of 
\cite{BCRR96, MT}.
Thus these arguments prove the existence of such fields on any maximal PN
manifold.

\subsection{Back to Bsq$\mathbf{5}$}
By computing the  minimal polynomial of $N^*$ of Eq.~\rref{eq:N} one
gets the $\boldsymbol{c}$'s coordinates
\begin{equation} \label{eq:cN}
\begin{array}{l}
c_1=k_1+h_2,\qquad
c_2=2k_1^2-h_2^2-g_1+2h_1 k_2-h_2k_1,\\
c_3=-3k_1^3+2k_1h_2^2+2k_1g_1-2h_1k_1k_2-2h_1h_2k_2+2k_2h_4-k_3k_2-h_2k_1^2,\\
c_4=k_2^2h_3-g_1k_1^2+h_2k_1^3-h_1^2k_2^2+k_1^4+2h_2h_1k_1k_2
-2k_1h_2h_4+k_1k_2k_3-h_2^2k_1^2 \ .
\end{array}
\end{equation}

The $\boldsymbol{f}$'s coordinates
can be found  as follows.  Taking into account the homogeneity properties of
the theory  with respect to the weights
$[\lambda]=3, \, [h_i]=1+i,\,  [k_i]=2+i,$
we see that $[P_0]=-8,\mbox{ and } [c_i]=3i$.  From~\rref{eq:ABCD} we get that 
the weight of $f_1$ must be equal to $-4$. Making the ansatz
$f_1=\dsl{\frac{1}{k_2}}$ one finds, using the recursion relation~\rref {f:rec} 
that the $\mathbf f$'s  are given by:
\begin{equation} \label{eq:fN}\begin{split}
f_1&=\frac{1}{k_2} \ , f_2=\frac{-h_2}{k_2} \ ,
f_3=-h_1+\frac{g_1+h_2^2-2k_1^2}{k_2} \\
f_4&=k_3-h_4+h_1 h_2+\frac{k_1^3+h_2k_1^2-k_1h_2^2-g_1k_1}{k_2}
\end{split}
\end{equation}
The Hankel--Fr\"obenius coordinates
associated with the PN structure allow us to reconstruct
the spectral curve \rref{spec_cur}, and the
Sklyanin separation equations~\cite{Sk} as follows.
In analogy with the corresponding arguments known from the KdV
case~\cite{Alber,DKN}, we consider equations~\rref{mula:f} as defining two
fundamental polynomials associated with the problem:
\begin{equation}
\CQ_1=\lambda^4-\sum_{i=1}^4 c_i \lambda^{(4-i)};\quad \CQ_2=\sum_{i=1}^4
f_i \lambda^{(4-i)};
\end{equation}
Then, a direct computation shows that on the zeroes $\lambda_j$ of
$\CQ_1$ and the corresponding values $\mu_j=\CQ_2(\lambda_j)$, the equation
of the spectral curve~\rref{spec_cur} is satisfied:
\begin{equation}
\mu_j^3=\mu_j(H_1
\lambda_j+H_2)+
(\lambda_j^5+H_3\lambda_j^3+H_4\lambda_j^2+H_5\lambda_j +H_6).
\end{equation}
When we restrict the problem to the leaf $S_0$, we have to consider $H_1=v_1$ and
$H_3=v_2$ as fixed {\em parameters}. Hence, we obtain
Sklyanin's separation equation~\cite{Sk}, i.e.
$4$ relations in the canonical coordinates $(\boldsymbol{\lambda};
\boldsymbol{\mu})$, depending on the $4$ commuting Hamiltonians 
$\widehat{H}_2,\widehat{H}_4,\widehat{H}_5,\widehat{H}_6.$
As it is well known, they
imply that, in analogy with the classical St\"ackel theory of
separation of variables for Hamiltonians depending quadratically on the
momenta, a complete
integral $W=W_1+W_2+W_3+W_4$ of the Hamilton--Jacobi equations
\[
\widehat{H}_i(\boldsymbol{\lambda},\frac{\partial W}{\partial
\boldsymbol{\lambda}})=\al_i,\quad i=2,4,5,6,
\]
can be gotten  by solving the four decoupled first--order ODE
\begin{equation}
\left(\frac{dW_j}{d\lambda_j}\right)^3=\frac{dW_j}{d\lambda_j}
(v_1\lambda_j+\al_2)+(\lambda_j^5+v_2\lambda_j^3+\al_4\lambda_j^2+\al_5\lambda_j
+\al_6)
\qquad j=1,\cdots, 4 \ ,
\end{equation}
a problem which can be treated by means of algebro--geometrical
techniques.

\section{Nijenhuis chains and Levi--Civita separability} \label{sec5}
Here we introduce the concept of Nijenhuis chains as a natural
extension of the Lenard sequences and provide and alternative and somehow more
classical proof of the fact that the Darboux--Nijenhuis coordinates  are
separation variables for the Hamilton-Jacobi equation corresponding
to each  Hamiltonian vector field $\widehat{X}_j$ of our problem.
Remark that, when dealing with
algebro--geometrical SoV in Section~\ref{sec4}, the original 
Hamiltonians~\rref{GZham}
of the ten--dimensional ``Gel'fand--Zakharevich'' problem of 
Proposition~\ref{prop:GZ} 
were not involved in the construction of the DN and HF coordinates, and
entered the separability proof only at the last stage.
The reason was that they, not being invariant under the
distribution $\widehat{\CD}$, cannot define a Lenard recursion relation on
$S_0$,
at least in the classical form enlightened, e.g., in~\cite{MM,Maetal}.
Here, we will consider their restriction $\widehat{H}_j$ to the symplectic
leaf $S_0$ in connection with  the following
criterion for separability due to Benenti~\cite{Benenti}:
\begin{prop} \label{pr:Benenti}
Let $X=P_0 dK$ a Hamiltonian vector field on a symplectic manifold $(M,
\omega=P_0^{-1})$ and
$(\mbox{\boldmath  $q$};\mbox{\boldmath $ p$)}$
 ($ \mbox{\boldmath $q$}:=(q_1,\ldots,q_n)$;
$\mbox{\boldmath $p$}:=(p_1,\ldots,p_n)$) a set of Darboux coordinates.
These coordinates are separation variables for the 
corresponding Hamilton-Jacobi
equation iff  $X$ admits $n$ integrals of motion $\{K_i\}_{1\le i \le n}$,
such that
\begin{equation}
\mbox{det}\left(\frac{\del K_i}{\del p_j}\right)\neq 0 \ ,
\end{equation}
and which are in
{\em separable} involution w.r.t. the chart $(\mbox{\boldmath
$q$};\mbox{\boldmath $
p$)}$, i.e. satisfy the $n$ relations
\begin{equation} \label{eq:sepinv}
\{K_i, K_j\}_{|k}:=
\frac{\partial K_i}{\partial q_k}\frac{\partial K_j}{\partial p_k}-
\frac{\partial K_i}{\partial p_k}\frac{\partial K_j}{\partial q_k}=0 \ ,
\end{equation}
where {\em no summation over $k$} is performed.
\end{prop}

In order to exploit this criterion, we introduce 
the notion of {\em Nijenhuis chain} which generalizes
the classical Lenard recurrence satisfied by the (normalized)
traces of powers of the Nijenhuis tensor $I_k={\frac{1}{2k}} \mbox{Tr}N^k$,
\[
N^* d I_k=d I_{k+1} \Rightarrow (N^*)^{k-1} dI_1=dI_k
\]
and the ones satisfied by the
HF coordinates $(\mbox{\boldmath  $f$};\mbox{\boldmath $ c$)}$ as a
consequence of (\ref{c:rec},\ref{f:rec}).

\begin{defi} \label{def:Nc}
Let $(M ,P,N)$ a PN manifold  and $\{K_j\}_{1\le j
\le n}$
$n$ independent functions which enjoy the
following property w.r.t. $N^*$
\begin{equation} \label{eq:Nh}
dK_j=\sum_{k=1}^n a_{jk}\,  (N^*)^{k-1} dK_1  \qquad  j=1,\cdots, n \ ,
\end{equation}
where $a_{jk}$ are the entries of an invertible matrix-valued function
$\boldsymbol{a}$. 
This recurrence relation (and the corresponding one  $X_j=P dK_j$ on the
vector fields) will be referred to as a
{\em Nijenhuis chain with  generator $K_1$ ($X_1=P dK_1$).}
\end{defi}
Nijenhuis chains were implicitly considered in~\cite{GT1} where it was   
proved that the functions  $K_j$ admitting the 
representation \rref{eq:Nh} are in involution w.r.t. the Poisson tensor $P$. 
Here, we want to state
a stronger property of such functions, namely that they are in
{\it separable} involution w.r.t. the DN coordinates.
This can be easily proved, by  computing
$\{K_i, K_j\}_{|k}$ in the DN coordinates, and recalling that
the $\lambda_j$ are eigenvalues of $N^*$.
Therefore in virtue of Proposition~\ref{pr:Benenti} we can state: 
\begin{prop} \label{pr:Nc}
The Darboux--Nijenhuis coordinates are separation variables for each
function $K_j$ belonging to a Nijenhuis chain.
\end{prop}
To prove that Bsq$_5$ fulfills Benenti's theorem,
it is thus sufficient to verify that the restrictions $\widehat{H}_j$, 
of the Hamiltonians $H_j$
to a generic symplectic leaf
$S_0$,  belong to a Nijenhuis chain.
One can directly check that this is  the case:
indeed, choosing as generator $K_1$ the Hamiltonian 
\[\begin{split}
\widehat{H}_2=&
{k_{{3}}}^{2}-h_{{4}}k_{{3}}+h_{{3}}{k_{{1}}}^{2}+{h_{{2}}}^{2}h_{{3}}
-2\,h_{{1}}{k_{{2}}}^{2}-3\,{k_{{1}}}^{2}k_{{2}}+{h_{{4}}}^{2}-h_{{1}}
k_{{1}}k_{{3}}+h_{{2}}h_{{1}}k_{{3}}+h_{{1}}k_{{2}}h_{{3}}\\ & -2\,h_{{1}}h
_{{2}}h_{{4}}-k_{{1}}k_{{2}}h_{{2}}-2\,k_{{1}}h_{{2}}h_{{3}}+2\,k_{{1}
}h_{{1}}h_{{4}}-k_{{1}}v_{{2}}+k_{{2}}g_{{1}}-k_{{2}}v_{{1}}+h_{{3}}g_
{{1}}-{h_{{1}}}^{2}g_{{1}}
\end{split},
\]
upon the identifications
$K_2=\widehat{H}_4,K_3=\widehat{H}_5,K_4=\widehat{H}_6$,
the matrix $\boldsymbol{a}$ of the Nijhenhuis recurrence 
is given (in the coordinates $(\boldsymbol{f}; \boldsymbol{c})$) by
\begin{small}
\begin{equation}
\boldsymbol{a}=
\left[
\begin {array}{cccc}
 1&0&0&0  \\  
-f_1 c_1-f_2 &f_1&0&0\\  
-f_1 c_2-f_3 &-f_1 c_1&f_1&0 \\ 
-f_1 c_3-f_4&-c_2 f_1&-f_1c_1&f_1
\end {array}\right ]
\end{equation}
\end{small}
\section{Summary}
In this paper we have discussed, on the ground of the concrete example
of the $t_5$ stationary Boussinesq system two related items.
In the first part of the paper we have sketched a
way to obtain stationary reductions, 
and their Hamiltonian structures, of
the $n$ Gel'fand--Dickey theories in a systematic way.
The key points were the study of the Central System~\rref{CS} and of its
stationary manifolds, and the zero--curvature
and Lax representation~\rref{ZS} one naturally gets for
the resulting flows.
Hamiltonian structures are induced on the stationary manifolds by means
of a specific bihamiltonian reduction procedure for a pair of Poisson
structures, defined on the space of Lie algebra--valued polynomials in the
spectral parameter $\lambda$.
This procedure, whose justification rests on the link between Lax and Poisson
formulation for Hamiltonian flows, lead us to frame such theories as the
Gel'fand--Zakharevich theories of a degenerate pencil of Poisson structures,
generically (i.e. for $n>2$ {\em  of non maximal rank} (see
Section~\ref{sec3})).

In the second part of the paper (Section~\ref{sec4} onwards) we addressed
the issue of how one can solve these integrable dynamical systems by the
method of separation of
variables. Our main theme was to study some of the relations between
the classical Hamilton--Jacobi theory (and also in its more recent aspects
related to the notion of complete {\em algebraic integrability})
and the Hamiltonian structure we got in the first part.
To this end, we had to abandon the GZ theory, and, by means of a further
(but {\em no more \bih\ }) reduction process, make contact with the theory of
Poisson--Nijhenhuis manifolds.
We studied the properties of two remarkable
classes of coordinate systems associated with the PN structures: the
Darboux--Nijhenuis coordinates and the Hankel--Fr\"obenius coordinates. By
using these
coordinates, we proved the separability of the Bsq$_5$ flows in two ways.
In the first we  made implicitly contact with the  ``method
of the poles of the Baker--Akhiezer
function'' of the Russian School (see, e. g.~\cite{DKN,Sk}).
In the second, via the introduction of the notion of Nijenhuis chain, we
proved separability by using an equivalent condition to the classical
Levi--Civita separability criterion~\cite{LC}.

As a final comment, we notice that
the peculiarity of the second reduction, which still
falls in the class of the MR reductions, is
given by the fact that  two different geometric processes are
used simultaneously:  the restriction for the
vector fields and the  projection for the bihamiltonian structure.
Due to this fact,  we were able to retain
the bihamiltonian structure (in the form of a PN structure) but lost
the bihamiltonian formulation for the vector fields.
This is not a new situation in the Hamiltonian theory of
integrable system and  happens, for instance, for:
\begin{itemize}
\item
the integrable H\'enon--Heiles system and its multidimensional generalizations
obtained by reduction from the {\it stationary} flows of the KdV hierarchy
\cite{GT1,GT2};
\item
a large class of  potentials recovered from the {\it
restricted} flows of the coupled
KdV systems, whose most representative member
is the Garnier system \cite{Bll}.
\end{itemize}
Such classes of dynamical systems live on a  bihamiltonian
manifold $M$ of maximal rank, and their Nijhenuis formulation,
namely a special case called quasi--bihamiltonian
formulation~\cite{BCRR96,MTPol}  can be gotten after a similar reduction.
The stationary flows  of the Boussinesq hierarchy
(and of the higher $n$ GD theories)
are  just examples of bihamiltonian
structures of non--maximal rank, whose interest may lie in the fact that an
analysis  similar to the one of~\cite{GZ93} for Poisson pencils of maximal rank
is not, to the best of our knowledge, at present available in the literature.
\subsection*{Acknowledgments}
We thank M. Pedroni and P. Casati for useful discussions.
G.F. and G.T. want to thank the organizers of {\em NEEDS in Leeds 1998}
for having given them
the opportunity to present these results there. Most computations have
been performed with Maple V$^{\circledR}$.

\end{document}